\documentclass[aps,prl,twocolumn,showpacs]{revtex4}
\usepackage{amsmath}
\usepackage{epsfig}
\usepackage{amssymb}
\usepackage{dcolumn}
\usepackage{graphicx}
\usepackage{dcolumn}

\begin{document}
\date{\today}

\title{A nonlinear quantum piston for the controlled generation of vortex rings}

\author{Natalia G. Berloff}
\email{N.G.Berloff@damtp.cam.ac.uk}

\affiliation{Department of Applied Mathematics and 
Theoretical Physics, University of Cambridge, United Kingdom.}

\author{V\'{\i}ctor  M. P\'erez-Garc\'{\i}a}
\email{victor.perezgarcia@uclm.es}

\affiliation{Departamento de Matem\'aticas, Escuela T\'ecnica
Superior de Ingenieros Industriales, and Instituto de Matem\'atica Aplicada a la Ciencia y la Ingenier\'{\i}a (IMACI),
Universidad de Castilla-La Mancha, 13071 Ciudad Real, Spain.}

\begin{abstract}
We propose a simple way to manage interactions in Bose-Einstein condensates to generate vortex rings in a highly controllable way.
 The vortex rings are generated under the action of a quantum analogue of a classical piston pushing the condensed atoms through a small aperture.
\end{abstract}

\pacs{03.75.Lm}

\maketitle

\emph{Introduction.-} One of the most remarkable achievements in quantum physics in the last decade was that of Bose-Einstein condensation in ultracold alkaline atomic gases. These physical systems have a high potential for supporting quantum nonlinear coherent excitations and in fact, many types of nonlinear waves have been experimentally observed \cite{dark,bright,gap,vector,v1,vrings,vrings2,vrings3,vrings5,sh1,blowup} or theoretically proposed to exist (see e.g. the reviews \cite{Panos})) in ultracold quantum degenerate gases, mainly with Bose-Einstein condensates (BECs).
In this way BECs are, apart from their fundamental interest, one of the best physical systems for the manifestation of nonlinear phenomena and certainly the best one in the quantum world.

But in the jungle of nonlinear excitations the king is probably the vortex ring. Vortex rings are essentially three-dimensional topological nontrivial structures appearing in either classical \cite{Batchelor} or quantum \cite{Donnelly} fluids. They are able to propagate in cylindrically trapped BECs as stable objects \cite{Komineas}, similarly to what happens in classical superfluids \cite{JR}. This is an essential difference with most other solitonic structures that become unstable when passing to fully three-dimensional scenarios, e.g. one dimensional bright solitons that are unstable to blow-up \cite{blowup} or dark solitons, that are unstable to the snake instability \cite{vrings}.  This leaves the vortex ring as the only dynamically nontrivial nonlinear excitation observed in BECs in three-dimensional scenarios.

Vortex rings were first observed in BECs as the outcome of the decay of dark solitons \cite{vrings} and as a result of the decay of quantum shock waves \cite{vrings2}. More recently they have been  observed to appear  during complex oscillations in  soliton-vortex ring structures \cite{vrings3} and during the merging of BEC condensate fragments \cite{Scherer}. However, all of those scenarios involve complicated nonlinear phenomena and in general
 a simple mechanism allowing for the controlled generation of a prescribed finite number of vortex rings is still missing. The purpose of this paper is to propose a mechanism allowing for the generation of a few vortex rings in a highly controllable way.

\emph{Physical idea.-} The process of vortex ring generation in classical fluids has received a substantial treatment in the literature. 
One of the most standard ways to obtain vortex rings in classical fluids involves moving a piston through a tube, resulting in a vortex ring being generated at the tube exit. A standard generation geometry consists of the tube exit mounted flush with a wall with the piston stroke ending at the tube exit \cite{Glezer}.

Is it possible to export this idea to the quantum world in order to obtain vortex rings in BECs? One might try to imagine a complicated setup to create a tube-like geometry using ordinary potentials and having a moving part in order to generate the flow, but this would lead to a complicated trap geometry and would be difficult to obtain experimentally. In this paper we will use something conceptually much simpler using the possibilities open by space-dependent Feschbach resonance management in a Bose-Einstein condensate. Since the first achievements in scattering length control in BECs \cite{FB1}, the technique of Feschbach resonance management has been improved and used in many different applications. Presently, the level of control of the scattering length allows for its very precise tuning \cite{Hulet} and nothing prevents an extended control of the interactions leading to a space dependent scattering length. A large number of theoretical papers have recently studied nonlinear phenomena in systems with managed interactions   \cite{Panos,V1,V2,V3}. 

The physical idea is very simple. Starting from an equilibrium BEC in a trap with a given value of the scattering length $a_1$ we propose modifying interactions in half of the space (say $z<0$) to a larger value $a_2>a_1$. This change would affect the initial configuration by inducing the transverse expansion of the atom cloud for $z<0$ while at the same time generating a flow towards the regions with smaller interaction values located 
at $z>0$. Those combined effects generate around $z=0$ a flow towards $z>0$
analogous to the piston-driven flow through an aperture used to generate vortex rings in classical fluids. It is interesting to note that all those effects are achieved simultaneously with a single action on the interactions without restorting to complicated external potentials. In what follows we will show that this idea leads to a highly controllable method for the generation of a small number of vortex rings. 

\begin{figure}
\epsfig{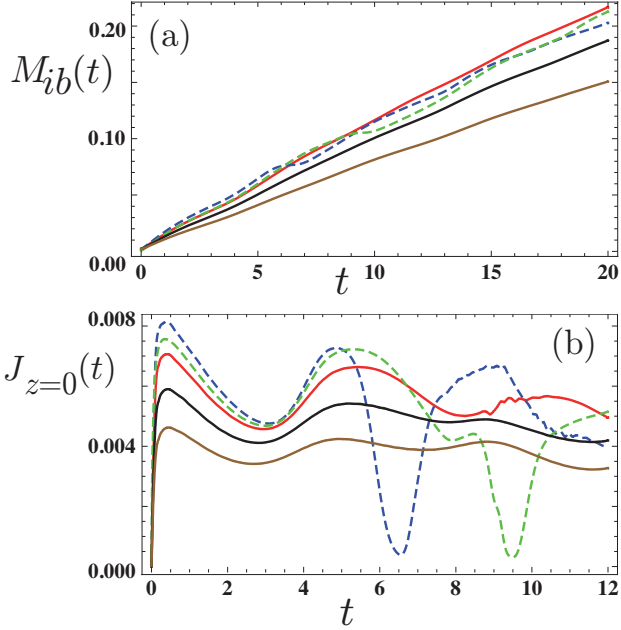}
\caption{Atom flow when increasing interactions for $z<0$ $(g_2)$ in an equilibrium condensate with $g_1 = 10^5, \lambda_x=\lambda_y = 1, \lambda_z = 0.05$ according to Eq. (\ref{model}). Shown are curves for $g_2/g_1=$1.6 (solid brown), 1.8 (solid black), 2 (solid red), 2.1 (dashed green), 2.2 (dashed blue).  Shown are (a) the mass imbalance $M_{ib}(t) = \int_{z>0} |\psi(x,y,z,t)|^2 dxdydz -  \int_{z>0} |\psi(x,y,z,t)|^2 dxdydz$ as a function of time and (b) the flow $J_{z=0}(t) = \int_S \rho u dS$ where $S$ is a surface orthogonal to the unit vector $\mathbf{u} = (0,0,u)$ at $z=0$. \label{prima}}
\end{figure}

\emph{Mathematical model.-} We will consider a BEC in the mean field limit ruled by the mean-field equations 
\begin{equation}\label{model}
2 i \frac{\partial \psi}{\partial t} = - \triangle \psi + V_{\text{ext}}(\boldsymbol{x})\psi + g(z) |\psi|^2 \psi,
\end{equation} 
where the spatial coordinates and time $t$ are measured in
units of $a_0=\sqrt{\hbar/2m \omega}$ and $1/2\omega$, respectively, while the energies and
frequencies are measured in units of $2\hbar \omega$ and $2\omega$ respectively, $\omega$ being the transverse
 frequency of the potential in physical units and $V_{\text{ext}} = (\lambda_x^2 x^2+\lambda_y^2 y^2+\lambda_z^2z^2)/2$. We assume that the total density is normalized to 1, so $\int |\psi|^2 dV = 1$. Also, $g(z) =4\pi a_s(z)/a_0$ is
proportional to the local value of the s-wave scattering length $a_s(z)$ and will be taken, starting from $t=0^+$, as
\begin{equation}
g(z) = \begin{cases} g_1, & z>0 \\ g_2, & z<0, \end{cases}\qquad g_2> g_1.
\label{g}
\end{equation}

\emph{Analytical estimates.-} The hydrodynamical form of Eq. (\ref{model})  for density $\rho$ and phase $\phi$ ($\psi = \rho^{1/2}e^{i\phi}$) consists of the continuity equation $\rho_t+\nabla\cdot (\rho {\bf u})=0$ and the integrated Bernoulli-type  equation $2\phi_t +  u^2 +  V_{\rm ext}+  g(z) \rho = \nabla^2 \sqrt{\rho}/\sqrt{\rho}$, where velocity is given by ${\bf u}=\nabla \phi$. These equations can be obtained from variational principle $\delta \int\int L dV dt=0$ for the Lagrangian $L=\rho (\partial_{x_i} \phi)^2/2+(\partial_{x_i}\rho)^2/8\rho + \rho V_{\text{ext}}+g(z)\rho^2/4$ leading to the momentum equation $\partial_t(\rho u_i) + \partial_{x_j} \Sigma_{ij}=0$, where the momentum flux density tensor $\Sigma_{ij}$ is given by $\Sigma_{ij} = (p+\rho \partial_{x_i} V_{\text{ext}})\delta_{ij} + \rho u_i u_j - \rho \partial^2_{{x_i}{x_j}} \ln \rho/4$ and the pressure $p=g(z)\rho^2/4$. The local speed of sound is given by $c^2=dp/2d\rho=g(z)\rho/2$.
 If $t<0$, then $g(z)=g_1$, and the system is in its ground state characterised by the  chemical potential $\mu$ such that $i\partial_t \psi=\mu \psi$. For $t<0$ in the Thomas-Fermi approximation we have $\mu_0=(15 g_1 \lambda_1 \lambda_2 \lambda_3/128 \pi)^{2/5}$ and density  $\rho_0=(2 \mu_0 - V_{\text{ext}})/g_1$. As $t\rightarrow \infty$ the system will reach a new ground state with $\mu_\infty=(15 g_1  g_2 \lambda_1 \lambda_2 \lambda_3/64 \pi(g_1+g_2))^{2/5}$, so that the density will become
\begin{equation}
\rho_\infty= \begin{cases} (2 \mu_\infty -V_{\text{ext}})/g_1 , & z>0 \\(2 \mu_\infty -V_{\text{ext}})/ g_2, & z<0, \end{cases}
\label{rho}
\end{equation}
 with $g_1/(g_1+g_2)$ particles on the left  and $g_2/(g_1+g_2)$ particles on the right of the plane $z=0$. At time $t=0^+$ when the interaction strengths are turned to step function given by Eq. (\ref{g}), we  have equal number of particles on both sides of $z=0$ plane, so $(g_2-g_1)/2(g_1+g_2)$ particles have to flow from the left half-plane to the right half plane before the equilibrium is reached.

 We consider the equations of conservation of mass and momentum in the neighbourhood of $z=0$. We can rewrite the mass conservation in integral form as
\begin{equation}
\frac{d}{dt} \int_{z_2}^{z_1} \rho dz + [\rho u]_{z_2}^{z_1} =0.
\end{equation}
 Taking the limits $z_i \rightarrow 0$ gives $\rho_1 u_1 = \rho_2 u_2$, where index $1$ ($2$) characterizes velocities and densities at $z \rightarrow 0^+$ ($z \rightarrow 0^-$). Similarly from the conservation of momentum equation and dropping the anisotropic quantum stress tensor  we have $\rho_1 u_1^2 + g_1 \rho_1^2/4=\rho_2 u_2^2 + g_2 \rho_2^2/4$. Combining these two equations gives
\begin{equation}
u_1^2= \frac{\rho_2}{4 \rho_1} \frac{g_1 \rho_1^2 - g_2 \rho_2^2}{\rho_1-\rho_2}, \quad u_2^2= \frac{\rho_1}{4 \rho_2} \frac{g_1 \rho_1^2 - g_2 \rho_2^2}{\rho_1-\rho_2}.
\label{u2}
 \end{equation}
Let's assume that  $\xi=g_2/g_1-1$ is small, so that $\delta=1-\rho_1/\rho_2$ is also small. The requirement $u_i \rightarrow 0$ as $\xi \rightarrow 0$ leads to $\delta = \xi/2$ and $u_2=\sqrt{g_1\rho_1\xi/2}$. From this it follows that $u_2$ (note that $u_2 > u_1$) may exceed the speed of sound $c$ if $\xi > 1$. So we expect the vortices to nucleate once $g_2/ g_1 > 2$. Our numerical simulations below confirm this prediction. 
 The rate of change of mass across the plane $z=0$ can than be estimated as $dM/dt=\int_{S_2} u_2\rho_2\, ds$ where $S_2$ is the cross-section at $z \rightarrow 0^-$. Approximating $\rho_i$ by $\rho_\infty$   gives 
\begin{equation}
\frac{dM}{dt}=\frac{3}{4}\lambda_z \frac{\sqrt{(g_2-g_1)g_1}}{(g_2+g_1)}.
\label{dm}
\end{equation}

\emph{Results.-} To test our ideas we have run many numerical simulations of our model equations (\ref{model}) for different parameter combinations. 
We have used a solver based on a fourth order finite difference scheme in space together with a fourth order Runge-Kutta discretization in time.

\begin{figure}
\epsfig{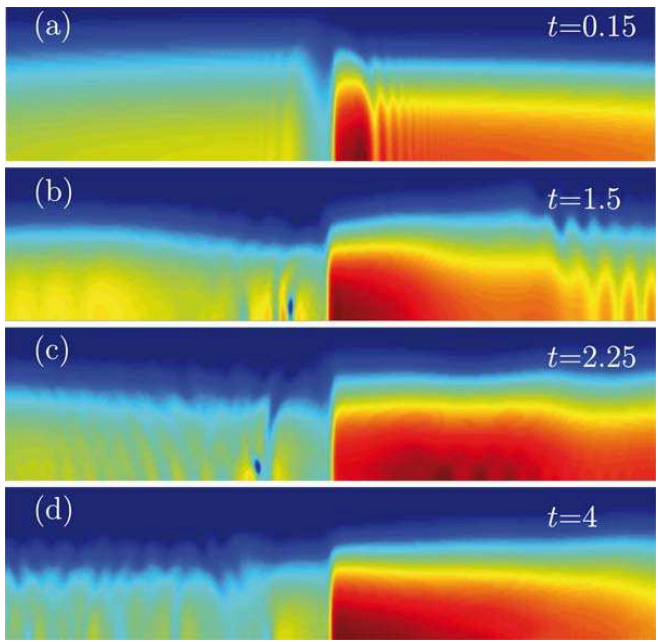}
\caption{(Color online) Pseudocolor plots of the atom density $|\psi(r,z,t)|^2$ for: (a) $t=0.15$, (b) $t=1.5$,  (c) $t=2.25$ and (d)  $t=4$. 
for $g_2/g_1 =2.1$. The spatial region shown corresponds to $z\in[-30,30], r\in [0,7]$. 
Blue corresponds to low atom densities and red to high ones. \label{ex1}}
\end{figure}

In what follows we describe the typical outcomes for a specific example corresponding to a large repulsive BEC with $g_1= 10^5$, with $\lambda_x=\lambda_y=1, \lambda_z = 0.05$ (i.e. soft longitudinal trapping) using a computational window of 240 space units in $z$ and $10$ units along the transverse directions $x,y$. Typical space steps are about 0.25 and the time steps $\Delta t \simeq 1.5\times 10^{-4}$. Our initial configuration is a ground state BEC corresponding to $g(z) = g_1$. On that initial configuration we suddenly increase interactions for $z<0$ and observe the subsequent condensate dynamics. 

\begin{figure}
\epsfig{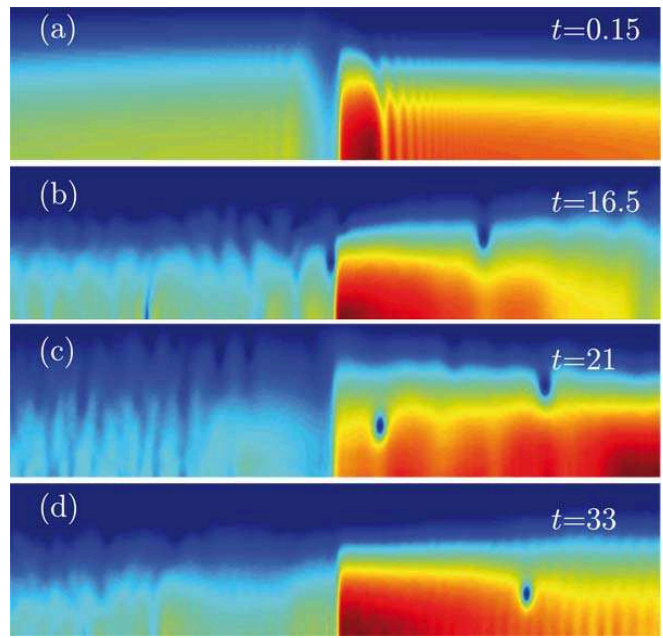}
\caption{(Color online) Same as in Fig. \ref{ex2} but for $g_2/g_1 =2.2$ and times (a) $t=0.15$, (b) $t=16.5$, (c) $t=21$, (d) $t=33$. \label{ex2}}
\end{figure}

\begin{figure}
\epsfig{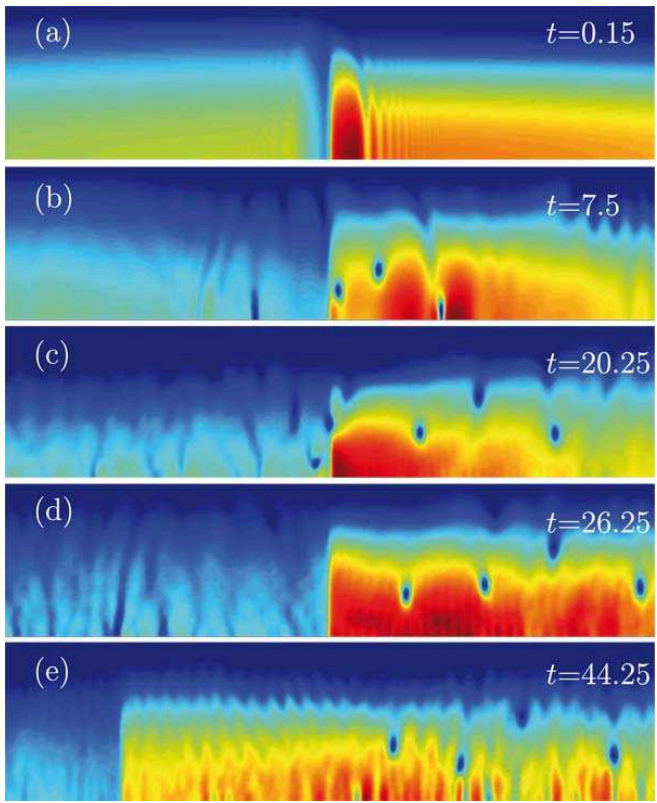}
\caption{(Color online) Pseudocolor plots of the atom density $|\psi(r,z,t)|^2$  for: (a) $t=0.15$, (b) $t=7.5$,  (c) $t=20.25$, (d)  $t=26.25$ and 
(e) $t=44.25$ for $g_2/g_1 =2.4$. The spatial region shown corresponds to $z\in[-30,30], r\in [0,7]$ in subplots (a-d) and to
 $z\in [-10,50]$ in subplot (e). Blue corresponds to low atom densities and red to high ones. \label{ex3}}
\end{figure}

The results of a series of simulations to be described in detail later are summarized in Fig. \ref{prima}. Once the nonequilibrium situation is generated there is a flow of atoms from $z<0$ to $z>0$ depending on the values of the ratio $g_2/g_1$. For ratios $g_2/g_1 \in[1,2]$ we find that our analytical predictions for mass rate of change and velocities of the flow are quite accurate below criticality (eq. from Eq. (\ref{dm}) we get $dM/dt\sim 0.01$ for $g_2 \sim 1.6$ as compared with Fig. \ref{ex1} ).  However, when a predicted critical value $g_2/g_1 \simeq 2$ is surpassed there are differences in the curves, such as a saturation in the mass transfer and a drastic reduction in the flow for certain times corresponding to the emission of vortex rings for those times as we will see later. Snapshots of the time evolution of the atom density are shown in Figs. \ref{ex1}-\ref{ex3}. 

First, in Fig. \ref{ex1} computed for $g_2/g_1 =2.1$  just around the critical value we observe that the initial spatial inhomogeneities in the interactions leads to a steady atom flow directed towards the region with smaller interactions. For small times the sudden increase of interactions results in a strong flow of atoms coming from the region with $z<0$ towards the region with $z>0$ (right) resulting in the formation of a shock wave \cite{V2}. Shortly after that, a small vortex ring appears in the low density regions around $z=0$ but it counterflows towards $z<0$ and disappears in that low density unstable region. After that, the situation remains stationary with time with some additional depletion of the condensate density for $z<0$ due to the remnant flow directed towards $z>0$.


In Fig. \ref{ex2} we present some results corresponding to $g_2/g_1 =2.2$ just above the critical value for vortex ring formation. After the initial shock wave [Fig. \ref{ex2}(a)] disappears from the central region, a vortex ring enters the condensate around $z=0$ coming from the low density regions corresponding to large $r$ values [Fig. \ref{ex2}(b)] and moves slowly through the condensate remaining stable for long times [Fig \ref{ex2}(c,d)]. Another vortex ring with a large radius seems to be present in the lower density regions where it would be experimentally difficult to detect.

Increasing the interactions even further to $g_2/g_1 =2.4$ leads to a richer dynamics as summarized in Fig. \ref{ex3}. The short-time dynamics is analogous to the previous cases [Fig. \ref{ex3}(a)] but then a complex transient appears where several vortex rings enter the condensate and also rarefaction pulses are clearly identified [see Fig. \ref{ex3}(b)]. After that, some of those vortices counter-flow and disappear and a very regular picture arises with several vortex rings moving to the right. Fig. \ref{ex3}(c) shows two vortex rings slowly moving through the condensate for $t=20.25$ and a third one being generated around $z=0$. Fig. \ref{ex3}(d) shows a later stage of the evolution where three long-lived vortex rings travel smoothly through the condensate although their relative positions change due to small differences in their speeds (notice the small differences in their radii) and their interaction with  sound waves originating after the reflection of the shock wave in the condensate boundary [see Fig. \ref{ex3}(e)].

\emph{Conclusions.-} We have proposed a simple method  to create a quantum piston able to generate vortex rings in Bose-Einstein condensates in a highly controllable way. Our proposal is accessible to present experimental techniques and improves essentially 
currently used methods to produce vortex rings that are based on the instability of nonlinear unstable structures and lead to a complex tangle of vortex lines and rings or to combined vortex rings and soliton complexes.

\acknowledgments

The authors acknowledge support from grants
No. EP/D032407/1 (EPSRC, UK), FIS2006-04190  (Ministerio de Ciencia e Innovaci\'on, Spain), and
PCI-08-093 (Junta de Comunidades de Castilla-La Mancha, Spain).

  \end{document}